\documentclass[10pt,letterpaper,twocolumn]{article} 

\usepackage{ol2}
\usepackage[draft,implicit=false]{hyperref}
\usepackage{amsmath}
\usepackage{epstopdf}

\begin{document}

\twocolumn[ 

\title{Focusing through random media in space and time: a transmission matrix approach}
\author{Zhou Shi,$^{1,2,*}$ Matthieu Davy,$^{1,3}$ Jing Wang,$^{1,2}$ and Azriel Z. Genack$^{1,2}$}
\address{
$^1$Department of Physics, Queens College of the City University of New York, \\Flushing, New York, 11367, USA\\
$^2$Graduate Center, City University of New York, New York, NY 10016, USA\\
$^3$Institut d'Electronique et des T\'{e}l\'{e}communications de Rennes, \\University of Rennes 1,Rennes, 35042, France\\
$^*$Corresponding author: zhou.shi@qc.cuny.edu
}
\begin{abstract}
We exploit the evolution in time of the transmission matrix following pulse excitation of a random medium to focus radiation at a selected time delay $t^\prime$ and position $r$. The temporal profile of a focused microwave pulse is the same as the incident Gaussian pulse. The contrast in space at time $t^\prime$ of the focused wave is determined by the participation number of transmission eigenvalues $M^\prime$ and the size $N^\prime$ of the measured transmission matrix. The initial rise and subsequent decay in contrast observed reflects the distribution of decay rates of the quasi-normal modes within the sample. 
\end{abstract}
] 

\noindent A short pulse incident upon a random medium is scrambled to produce a multiply peaked speckle pattern of intensity in space and time. The scattering of light severely compromises potential applications in imaging, two-photon microscopy, nano-surgery and telecommunications \cite{1,2,3,4}. The ability to focus a pulse after it is transmitted through a random medium was first demonstrated in acoustics by time reversing the transmitted signal \cite{5}. The transmitted pulse is picked up by an array of transducers and then played back in time. The signal converges to a focus at the location of the source antenna. Recently, optical pulses have also been focused through random samples by shaping the wavefront \cite{6} using a genetic learning algorithm based on feedback of light arriving at a selected point and time delay \cite{7} or via nonlinear feedback of two photon fluorescence \cite{8}.

Transmission through a random system is fully described by the field transmission matrix $t$, whose elements $t_{ba}$ are the transmission coefficients of the field between incoming and outgoing channels, $a$ and $b$, respectively \cite{9,10}. Measurements of $t$ have been used to focus monochromatic light \cite{11} and microwave radiation \cite{12} through random samples via phase conjugation \cite{11,12,13}, in which the electric field from different points $a$ on the incident plane arrive in phase at the target to interfere constructively at a maximum value of focused intensity. The spatial contrast $\mu$ between the average intensity at the focus and the average background intensity for monochromatic illumination is determined by the measured eigenchannel participation number of the transmission matrix, $M^\prime \equiv (\sum_{n=1}^{N^\prime} \tau_n)^2/\sum_{n=1}^{N^\prime} \tau_n^2$ \cite{14}. Here, $N^\prime$ is the size of the measured transmission matrix and $\tau_n$ are the eigenvalues of the matrix product $tt^\dagger$. 

In this Letter, we show that pulsed radiation can be focused in space and time by phase conjugating a time dependent microwave transmission matrix at a selected time $t^\prime$. The temporal profile of the focused pulse is the square of the field correlation function in time. This is independent of time and for a Gaussian incident pulse is equal to the profile of the incident pulse \cite{15}. The spatial contrast at $t^\prime$ is determined by the eigenchannel participation number $M^\prime(t^\prime)$ and size $N^\prime$ of the measured transmission matrix at that time, $M^\prime (t^\prime)=(\sum_{n} \tau_n(t^\prime))^2/\sum_{n} \tau_n^2(t^\prime)$, in which the sum of $\tau_n(t^\prime)$ yields the transmission at that time. We find that as soon as a random speckle pattern forms in transmission, the time evolution of the contrast in focusing equals $1/(1/M^\prime(t^\prime)-1/N^\prime)$. The contrast in focusing reaches a peak and then decays as a result of the time variation of the effective number of quasi-normal modes contributing to transmission. 

The samples studied were composed of alumina spheres with a diameter of 0.95 cm and refractive index 3.14 embedded in low-index Styrofoam shells. The randomly positioned spheres fill in a copper tube of diameter 7.3 cm at an alumina filling fraction of 0.068. The field transmission matrix $t$ is measured for microwave propagating through random waveguides with length $L= 61$ cm over the frequency range 14.7-15.7 GHz in 3200 steps. The number of propagating waveguide modes $N$ of the empty waveguide varies from 64 to 72 over this frequency range. The field transmission coefficients for radiation polarized along the detecting antenna is measured between all pairs of source and detection antennas positioned on a grid of $N^\prime$=45 points on the input and output surfaces of the sample tube. The grid spacing of 9 mm is close to the field correlation length at which the field correlation function passes through zero. New statistically equivalent samples are produced by momentarily rotating and vibrating the copper tube after $t$ is recorded. The sample studied is diffusive with an average value of the steady-state transmittance $\langle T\rangle \sim 4$. $\langle T\rangle$ is equal to the dimensionless conductance {\textsl g}. Weak absorption in the alumina spheres and in reflection from the copper tube reduces transmission by 25\%. The impact of dissipation upon the measured contrast is reflected in a reduction in $M^\prime$ in the steady state measurement of the order of 2\% in the samples studied.
\begin{figure}[htc]
\centering
\includegraphics[width=2.5in]{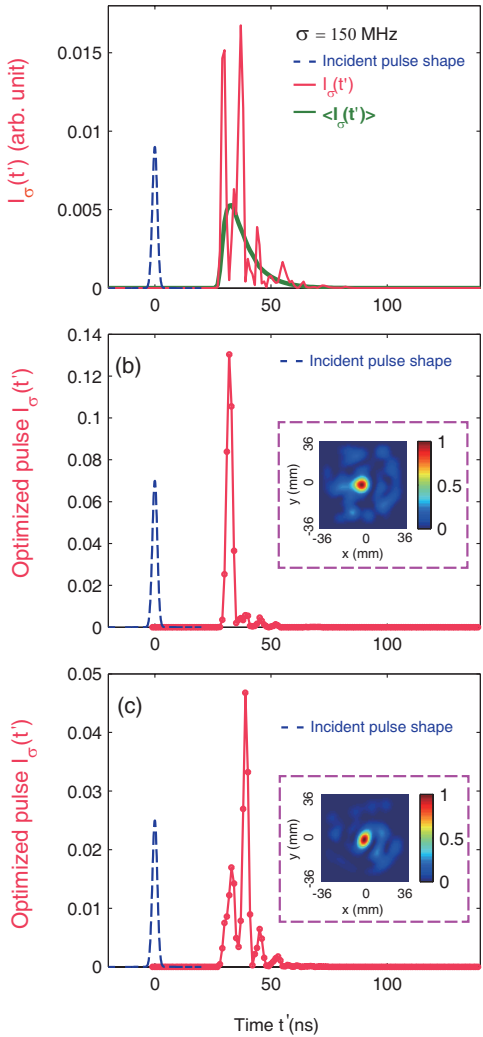}
\caption{(Color online) Spatiotemporal control of wave propagation through a random waveguide. (a) Typical resposne of $I_{ba}(t^\prime)$ and the time of flight distribution $\langle I(t^\prime)\rangle$ found by averaging over an ensemble of random samples. The incident pulse is sketched in the dashed blue curve. (b) and (c), phase conjugation is applied numerically to the same configuration as in (a) to focus at $t^\prime$= 33 ns and 40 ns at the center of the output surface in (b) and (c), respectively. The Whittaker-Shannon sampling theorem is used to obtain high-resolution spatial intensity patterns shown in the inset of (b) and (c).} \label{Fig1}
\end{figure}

We obtain the time dependent transmission matrix from spectra of the transmitted field between all points $a$ and $b$, $t_{ba}(\nu)$. These spectra are multiplied by a Gaussian pulse centered in the measured spectrum at $\nu_0=$15.2 GHz with bandwidth $\sigma_\nu=$150 MHz and then Fourier transformed into the time domain. This gives the time response at the detector to an incident Gaussian pulse launched by a source antenna with bandwidth $\sigma_t$=$1/2\pi/\sigma_\nu$. The time variation $I_{ba}(t^\prime)$ of an incident pulse launched at the center of the input surface and detected at the center of the output surface in a single realization of the random sample is shown in Fig. 1(a). Individual peaks in intensity have widths comparable to the width of the incident pulse. The average of the time of flight distribution $\langle I(t^\prime)\rangle$ over transmission coefficients for $N^{\prime 2}$ pairs of points and 8 random configurations is also presented and seen to be significantly broadened over the incident pulse. 

The intensity that would be delivered to a point at the center of the output surface of the waveguide $\beta$=(0,0) at a selected time $t^\prime$ if the transmission matrix were phase conjugated at time $t^\prime$ is investigated. The results for $t^\prime$= 33 ns and 40 ns are shown in Figs. 1(b) and 1(c), respectively. In both cases, a sharp pulse emerges at the selected time delay with intensity peaked at $\beta$=(0,0).
\begin{figure}[htc]
\centering
\includegraphics[width=3in]{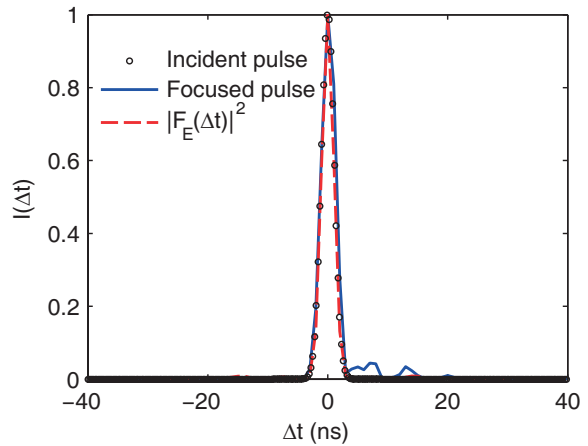}
\caption{(Color online) Profile of the focused pulse compared with the square modulus of the field correlation function in time and the profile of incident Gaussian pulse. The focused pulse in Fig. 1(b) has been plotted relative to the time of the peak. All curves are normalized to unity at $\Delta t$=0 ns.} \label{Fig2}
\end{figure}	

The spatial profile of focusing for a monochromatic wave above an enhanced constant background is equal to the square of the field correlation function in space \cite{12,16}. Similarly, the temporal profile of the focused pulse is seen in Fig. 2 to correspond to the square of the field correlation function in time, $|F_E^\sigma(\Delta t)|^2$, where $F_E^\sigma\equiv \langle E_\sigma(t^\prime)E_\sigma^*(t^\prime+\Delta t)\rangle/(\langle I(t^\prime)\rangle \langle I(t^\prime+\Delta t)\rangle)^{1/2}$. For an incident Gaussian pulse, the square of the field correlation function is equal to the intensity profile of the incident pulse and is independent of delay time.

We have shown previously that in a large single transmission matrix in steady state, $\mu=1/(1/M-1/N)$ \cite{14}. Here, $M$ is the eigenchannel participation number when the full transmission matrix of size {\it N} is measured. Because the full transmission matrix is not accessible in the experiment, the density of measured transmission eigenvalues differs from theoretical prediction \cite{11,17,18,19}. Nonetheless, we find in steady state measurements \cite{14} that when part of the transmission matrix is measured, the contrast in focusing via phase conjugation is given by, 
\begin{align}
\mu=1/(1/M^\prime-1/N^\prime).
\end{align}           
where $M^\prime$ is the eigenvalue participation number of the measured transmission matrix of size $N^\prime$. This is a property of random transmission matrices and therefore should apply as well to transmission matrices at different time delays provided the field within the transmission matrix is randomized. In Fig. 3, we present the time evolution of $\langle M^\prime\rangle$ and $\langle \mu\rangle$. Near the arrival time of the ballistic wave, the value of $M^\prime$ is close to unity and the contrast is not described by Eq. (1). This is because ballistic wave is associated with the propagating waveguides modes with the highest speed and therefore the transmitted field is not randomized. Once the transmitted waves at the output have been multiply scattered, a random speckle pattern develops and the measured contrast is in accord with Eq. (1). 
\begin{figure}[htc]
\centering
\includegraphics[width=3in]{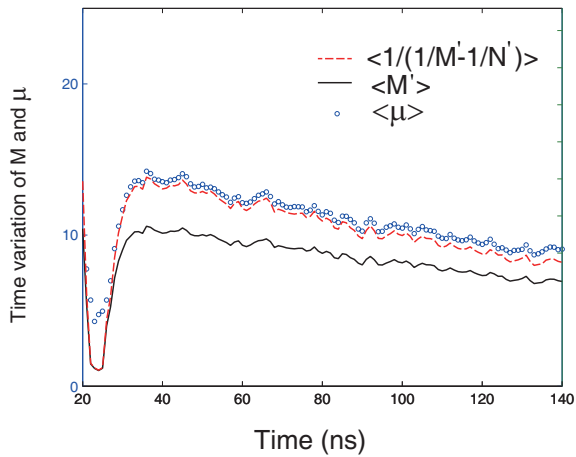}
\caption{(Color online) Time evolution of $\langle M^\prime \rangle$ (lower solid curve) and the maximal focusing contrast $\langle \mu \rangle$. $\mu$ is well described by Eq. (1) after the time of the ballistic arrival, $t^\prime \sim 21$ ns. At early times, the signal to noise ratio is too low to analyze the transmission matrix.} \label{Fig3}
\end{figure}

After the arrival of ballistic waves, the value of $M^\prime$ is seen in Fig. 3 to increase rapidly before falling slowly. This reflects the distribution of lifetimes and the degree of correlation in the speckle patterns of quasi-normal modes \cite{20,21,22}. Just after the ballistic pulse, transmission is dominated by the shortest-lived modes. These modes are especially short lived and strongly transmitting because they are extended across the sample as a result of coupling between resonant centers \cite{23,24,25}. Sets of extended modes that are close in frequency could be expected to have similar speckle patterns in transmission \cite{22}, so that a number of such modes might then contribute to a single transmission channel. As a result, the number of independent eigenchannels of the transmission matrix contributing substantially to transmission would be relatively small at early times and $M^\prime$ would be low. At late times, only the long-lived modes contribute appreciably to the transmission. Thus for intermediate times, modes with wider distribution of lifetimes than at either early or late times contribute to transmission and these modes are less strongly correlated than at early times so that $M$ and the contrast are peaked. 

In conclusion, we have shown that measurements of spectra of microwave field transmission coefficients between pairs of source and detector antenna positions make possible a detailed comparison of focusing calculations and measurements. These results provide the conditions for optimal focusing of light and sound at a selected time. The time profile of the focused pulse is equal to the square of the field correlation function in time. The spatial contrast at the time the wave is focused depend only upon $M^\prime$ and $N^\prime$ of the measured transmission matrix at that time. The time evolution of maximal focusing is related to the range of decay times of the underlying quasi-normal modes and the degree of correlation of the modal speckle patterns. This study provides a framework for focusing optical pulses for bio-medical and communications applications and for linear as well as nonlinear microscopies.

The research was supported by the NSF under Grant No. DMR-1207446 and by the Direction G\'{e}n\'{e}rale de l$^\prime$ Armement (DGA).

\end{document}